\documentclass[prl,twocolumn,superscriptaddress,showpacs,floatfix]{revtex4-1}

\usepackage{epsfig,graphicx,amsmath,amsfonts,amssymb,times}
\usepackage[sort&compress]{natbib} 
\begin{document}

\title{Correlations transference and apparition of a metastable decoherence-free subspace in dissipative reservoirs }

\date{\today}
	
\begin{abstract}
The dynamic of correlations in a system composed of a two-mode quantum field coupled with the environment is studied. The quantum field corresponds to two entangled coherent states whose amplitude we vary up to the mesoscopic regime. We show that under the onset of decoherence, correlations in the quantum field are not lost but transferred to the environment. We also found that sudden transitions in the decoherence regimes appear along with the dynamics depending on the coherent states' amplitude. Increasing the amplitude of the entangled coherence state results in the apparition of a metastable decoherence-free subspace (DFS) in the field subsystem, and the transference of classical correlations freezes. This subspace only exists during a time interval that depends on the average number of photons. Interestingly, the reservoir subsystem also experiences the apparition of a DFS. Only quantum correlations are transferred while the DFS exists. 

\end{abstract}

\author{F. Lastra}
\affiliation{Departamento de F\'{\i}sica, Facultad de Ciencias B\'{a}sicas, Universidad de Antofagasta,   
Casilla 170,  Antofagasta, Chile}

\author{C.E. L\'opez}
\affiliation{Departamento de F\'{\i}sica, Universidad de Santiago de
Chile, USACH, Casilla 307 Correo 2 Santiago, Chile}
\maketitle

Decoherence is the loss of a system's quantum behavior, which generates a transition from a quantum system to a classical one. A deep understanding of this process is key to both quantum theory and its applications. The decoherence have a variety of theoretical approaches~\cite{Zeh70,zurek81,Kupsh03,zurek03,zurek97,buhmann12,machnikowski06,howie11}.

A situation where the onset of decoherence is relevant describes an initially correlated bipartite system coupled to the environment. When the bipartite system is composed of two qubits (two-level systems), entanglement may suddenly disappear given its interaction with the environment~\cite{zy,dio,dod,yu,san}. However, entanglement is not the only correlation that may be present in this or other quantum systems. Moreover, applications such as quantum computing are possible even in the absence of entanglement~\cite{lan}. Such correlations are a central problem of quantum mechanics and a key to the development of quantum technologies. Therefore, the study of the effects of decoherence on these correlations is a significant problem addressed \cite{sha, dat, pia, wer, maz1}. All possible quantum and classical correlations in a bipartite system are described through quantum mutual information~\cite{quantummutual, oli}. From this, quantum correlations (known as discord) and classical correlations are distinguishable~\cite{oli, hen,opp,Luo1,mod}.

As we discussed previously, it is relevant to study decoherence on classical and quantum correlations. Interesting results appear when studying the decay of total correlations (quantum mutual information) and distinguishing how much of this decay is associated with classical correlations or quantum correlations. For example, when a two-qubit system evolves under dephasing, for a certain time, its classical correlations decay while the quantum ones remain constant. Following this, classical correlations freeze, and the quantum ones begin to decay \cite{maz, mazz}, as been observed for non-markovian and markovian reservoirs \cite{xu,cor}. Also, it has been shown that quantum correlations can be frozen for non dissipative dynamics \cite{maniscalco,aaronson,lastra2014,lastra17} leading to the apparition of a pointer basis \cite{oli,hen,cor,lastra2014,lastra17,lastra18,zur}.

In this Letter, we study the transfer of correlations from a bipartite system of mesoscopic quantum fields to the environment. We assume that these systems do not interact with each other but are coupled to independent reservoirs. In this situation, we show that both the information and the correlations initially present in the bipartite system are entirely transferred to the reservoirs. Previously, we have shown that several time scales appear during evolution. One of these time scales marks the appearance of a metastable decoherence-free subspace~\cite{lastra18}. Here, we demonstrate that the apparition of a mirror decoherence-free subspace in the reservoirs. These subspaces' duration will depend on each field's average number of excitations at $ t = 0 $.

Here, we consider a physical system composed of two non-interacting quantized harmonic oscillators modes, for example, two cavity QED modes. Each cavity mode is coupled to a dissipative reservoir. We denote the cavities' subsystem as $c_1 \otimes c_2$ and for reservoir's subsystem we use $r_1 \otimes r_2$. Given that each quantum mode interacts only with its own reservoir, the quantum dynamics can be obtained considering each cavity - reservoir subsystem $c_j \otimes r_j$,  separately. In this scenario, the Hamiltonian of a subsystem $c_j \otimes r_j$ can be written, in the interaction picture, as follows
\begin{equation}
\hat{H} =  \hbar\sum_k \left(g_k \hat{a} ^\dagger \hat{b}_k e^{i(\nu-\nu_k)t}+g_k ^*\hat{a} \hat{b}_k^\dagger e^{-i(\nu-\nu_k)t} \right),
\label{H1}
\end{equation}
where the operator $\hat{a}^\dagger (\hat{a})$ creates (annihilates) a photon in the cavity $c_j$ and $\hat{b}_k^\dagger (\hat{b}_k)$ creates (annihilates) an excitation in the $k$-th mode of the reservoir $r_j$.

To investigate the transference of classical and quantum correlations from the cavity modes to the reservoirs, we have to explicitly consider the reservoir's dynamics. That is, we cannot use typical techniques such as master equations that trace out reservoir modes. That is, we cannot use typical techniques such as master equations that trace out reservoir modes.  With this in mind, we first consider a single cavity mode in a coherent state $|\alpha\rangle_c$ and all reservoir modes in the vacuum state $|\bar{0}\rangle_r\equiv\bigotimes_{k} |0_k \rangle_r$. The unitary evolution given by Hamiltonian~(\ref{H1}) for the initial state $|\psi_0 \rangle_{c-r} = |\alpha\rangle_s \bigotimes_{k} |0_k \rangle_r $, can be written in the Born-Markov approximation as: 
\begin{equation}
|\psi_t \rangle_{c-r} = \hat{U} |\alpha\rangle_c \bigotimes_{k} |0_k \rangle_r \equiv |\alpha_t\rangle_c |\bar{\alpha}_t\rangle_r,
\label{ev1}
\end{equation}
where $\hat{U}=\exp{(-i \hat{H} t /\hbar) }$, $\alpha_t=\alpha e^{-\gamma t/2} $ and $|\bar{\alpha}\rangle_r =  \bigotimes_{k} |d_k \alpha\rangle_r$. Amplitudes $d_k$ satisfy that $\sum_k d_k^{2}=1-e^{-\gamma t}$. Once the single $c\otimes r$ system is solved, we now can  consider the following initial density matrix for the overall system:
\begin{equation}
\rho(0)= \rho_{c_1 c_2}\otimes |\bar{0}\bar{0}\rangle_{r_1 r_2} \langle \bar{0}\bar{0}|.
\label{rho0}
\end{equation}
Here, we consider that reservoirs $r_1$ and $r_2$ are initially in the vacuum state while cavity modes $c_1$ and $c_2$ are prepared in an incoherent superposition of entangled coherent states with mean number of excitations $\bar{n}=|\alpha|^2$ such that,
\begin{equation}
\rho_{c_1 c_2} = p |\psi\rangle\langle \psi| + (1-p)  |\phi\rangle\langle \phi|,\label{rhoc1c20}
\end{equation}
with
\begin{eqnarray*}
|\psi\rangle &=& \frac{1}{f_+ (\alpha)} \left(|\alpha\rangle_{c_1} |\alpha \rangle_{c_2} +|-\alpha\rangle_{c_1}|-\alpha\rangle_{c_2}\right), \\
|\phi\rangle &=&  \frac{1}{f_+ (\alpha)}\left(|\alpha\rangle_{c_1} |-\alpha \rangle_{c_2} +|-\alpha\rangle_{c_1}|\alpha\rangle_{c_2}\right),
\end{eqnarray*}
and $f_\pm^2 (x)=2(1 \pm e^{-4 x^2})$.

Considering Eq.~(\ref{ev1}) and the density matrix~(\ref{rho0}), finding $\rho(t) = \hat{U} \rho(0)\hat{U}^\dagger$ is straightforward. If we extend Hamiltonian~(\ref{H1}) to account for the interaction of both cavities with their corresponding reservoir, we can infer that dynamics will entangle the cavity $c_1$ with $r_1$ and cavity mode $c_2$ with the reservoir $r_2$. Moreover, given the subsystem $c_1 \otimes c_2$ is initially entangled, the subsystem $r_1 \otimes r_2$ will also become entangled.

To prove this, it is convenient to write the initial density matrix $\rho(t)$ in the time-dependent basis: $\{|\pm\rangle_c,|\tilde{\pm} \rangle_r \}$ where,
\begin{eqnarray}
|\pm\rangle_{c} &=& \frac{1}{g_\pm (\alpha_t)}(|\alpha_t\rangle \pm |-\alpha_t\rangle),\\
|\bar{\pm}\rangle_{r} &=& \frac{1}{g_\pm (\bar{\alpha}_t)}(|\bar{\alpha}_t\rangle \pm |-\bar{\alpha}_t\rangle),
\label{catbasis}
\end{eqnarray}
with $g_\pm^2 (x)=2(1 \pm e^{-2 x^2})$, $\alpha_t^2 = \bar{n}\exp{(-\gamma t)}$ and $\bar{\alpha}_t^2 =\bar{n}(1-\exp{(-\gamma t)})$. Written in this basis, all parties of the overall system $c_1 \otimes c_2 \otimes r_1 \otimes r_2$  are effective two-level systems (qubits). This will allow us to calculate  quantum and classical correlations in the bipartite systems $c_1 \otimes c_2$ and $r_1 \otimes r_2$. Thus, we first calculate the dynamics for the partition $c_1 \otimes c_2$ by tracing out the reservoir modes. In the basis $\{|++ \rangle_{c_1 c_2}, |+- \rangle_{c_1 c_2}, |-+ \rangle_{c_1 c_2}, |--\rangle_{c_1 c_2}\}$, the reduced density matrix for the cavities subsystem is given by
\begin{equation}
\begin{tabular}{l}
$\rho _{c_1 c_2}=\frac{1}{16 f_+^2(\alpha)}\left(
\begin{array}{cccc}
r_{11} & 0 & 0 & r_{14} \\0 & r_{22} & r_{23} & 0 \\
0 & r_{32} & r_{33} & 0 \\
r_{41} & 0 & 0 & r_{44}%
\end{array}%
\right) ,$%
\end{tabular}
\label{rhoc1c2}
\end{equation}%
with matrix elements:
\begin{eqnarray}
r_{11} &=& g^4_+(\alpha_t) f^2_+(\bar{\alpha}_t), \notag \\
r_{44} &=& g^4_-(\alpha_t) f^2_+(\bar{\alpha}_t), \notag\\
r_{22} &=& r_{33} = g^2_+(\alpha_t) g^2_-(\alpha_t) f^2_-(\bar{\alpha}_t), \\
r_{14} &=& r_{41} = \left(2p - 1\right) g^2_+(\alpha_t) g^2_-(\alpha_t) f^2_+(\bar{\alpha}_t), \notag\\
r_{23} &=& r_{32} = \left(2p - 1\right) g^2_+(\alpha_t) g^2_-(\alpha_t) f^2_-(\bar{\alpha}_t).\notag 
\label{coefs1s2}
\end{eqnarray}

On the other hand, to calculate the dynamics of the reservoirs partition, we need to trace out the cavities subsystem. It is not difficult to find that the reduced density matrix for the reservoirs subsystem $\rho_{r_1 r_2}$ has the same structure than $\rho_{c_1 c_2}$ in Eq.~(\ref{rhoc1c2}) where the matrix elements $\tilde{r}_{ij}$ are the ones in~(\ref{coefs1s2}) but exchanging $\alpha_t \leftrightarrow \bar{\alpha}_t$. For example,  $\tilde{r}_{11} = g^4_+(\bar{\alpha}_t) f^2_+(\alpha_t)$.

In the following, we use this results to address the question of how correlations are transferred from partition $c_1 \otimes c_2$ to partition $r_1 \otimes r_2$. A bipartite quantum system $\hat{\rho}_{{ab}}$ as the one described above, can feature both quantum and classical correlations. Total correlations are characterized by the quantum mutual information $I(\hat{\rho}_{{ab}})=S(\hat{\rho}_{{a}})+S(\hat{\rho}_{{b}})-S(\hat{\rho}_{{ab}})$,
where $S(\hat{\rho})=-{\rm Tr}[\hat{\rho}\log_2(\hat{\rho})]$ is the von Neumann entropy. Based on this expression, 
correlations can be separated according to their classical and quantum nature, respectively. In this way the quantum discord
has been introduced as $D_{ab}=I(\hat{\rho}_{ab})-C(\hat{\rho}_{ab})$ which quantifies genuine quantum correlations, including correlations that can be distinct from entanglement. Here $C(\hat{\rho}_{ab})$ are the classical correlations defined by \cite{oli,hen,opp,Luo1,mod} 
\begin{equation}
	C_{ab}=\max_{\{\hat{\Pi}_{k}\}}\left[S(\hat{\rho}_{a})-S(\hat{\rho}_{ab}\mid\{\hat{\Pi}_{k}\})\right]. \label{cc}
\end{equation}	 
The optimization is carried out with respect  all possible complete set of projector
operators  $\{\hat{\Pi}_{k}\}$ for the subsystem
$b$, and $S(\hat{\rho}_{ab}\mid\{\hat{\Pi}_{k}\})=\sum_{k}p_{k}S(\hat{\rho}_{k})$,
$p_{k}={\rm Tr}(\hat{\rho}_{ab}\hat{\Pi}_{k})$, and $\hat{\rho}_{k}={\rm Tr}_{b}(\hat{\Pi}_{k}\hat{\rho}_{ab}\hat{\Pi}_{k})/p_{k}$. This can be understood as the amount of information we can retrieve about one party (here, system $a$) by measuring the other one (system $b$). For matrices such Eq.~(\ref{rhoc1c2}) know as $X$-states, classical and quantum correlations can be solved analytically \cite{xstate,CHOh2011}. More precisely, when $\left(|r_{23}|+|r_{14}| \right)^2 \leqslant \left(r_{11}-r_{22}\right)\left(r_{44}-r_{33}\right)$ the optimal observables that maximize (\ref{cc}), corresponds to $\sigma_z$. Now if  $|\sqrt{r_{11} r_{44}}-\sqrt{r_{22} r_{33}}| \leqslant |r_{23}|+|r_{14}|$, the optimal observable is $\sigma_x$. In such case, the expression for the classical correlations are now given by

\begin{equation}
	C_{ab} =S(\hat{\rho}_{a})- \min_{\{\sigma_x,\sigma_z\}}\left[S(\hat{\rho}_{ab})\rvert \{\sigma_x,\sigma_z\}\right],\label{cca}
\end{equation} 
where $S(\hat{\rho}_{ab})\rvert \{\sigma_x,\sigma_z\}$ is the von Neumann entropy of $\hat{\rho}_{ab}$ when $\sigma_x$ or $\sigma_z$ has been measured in the subsystem $B$. When the classical correlation $C_{ab}$ is maximized by measuring the observable $\sigma_z $, we say that: $ C_ {ab} = C^{Z}_{ab} \equiv S(\hat{\rho}_{a}) - S(\hat{\rho}_{ab})\rvert \{\sigma_z\} $. Now, if $C_ {ab}$ is maximized when we measure $\sigma_x$, we say that $C_ {ab} = C^{X}_{ab} \equiv S(\hat{\rho}_{a}) - S(\hat{\rho}_{ab})\rvert \{\sigma_x\}$.

\begin{figure}[t]
	\includegraphics[width=85mm]{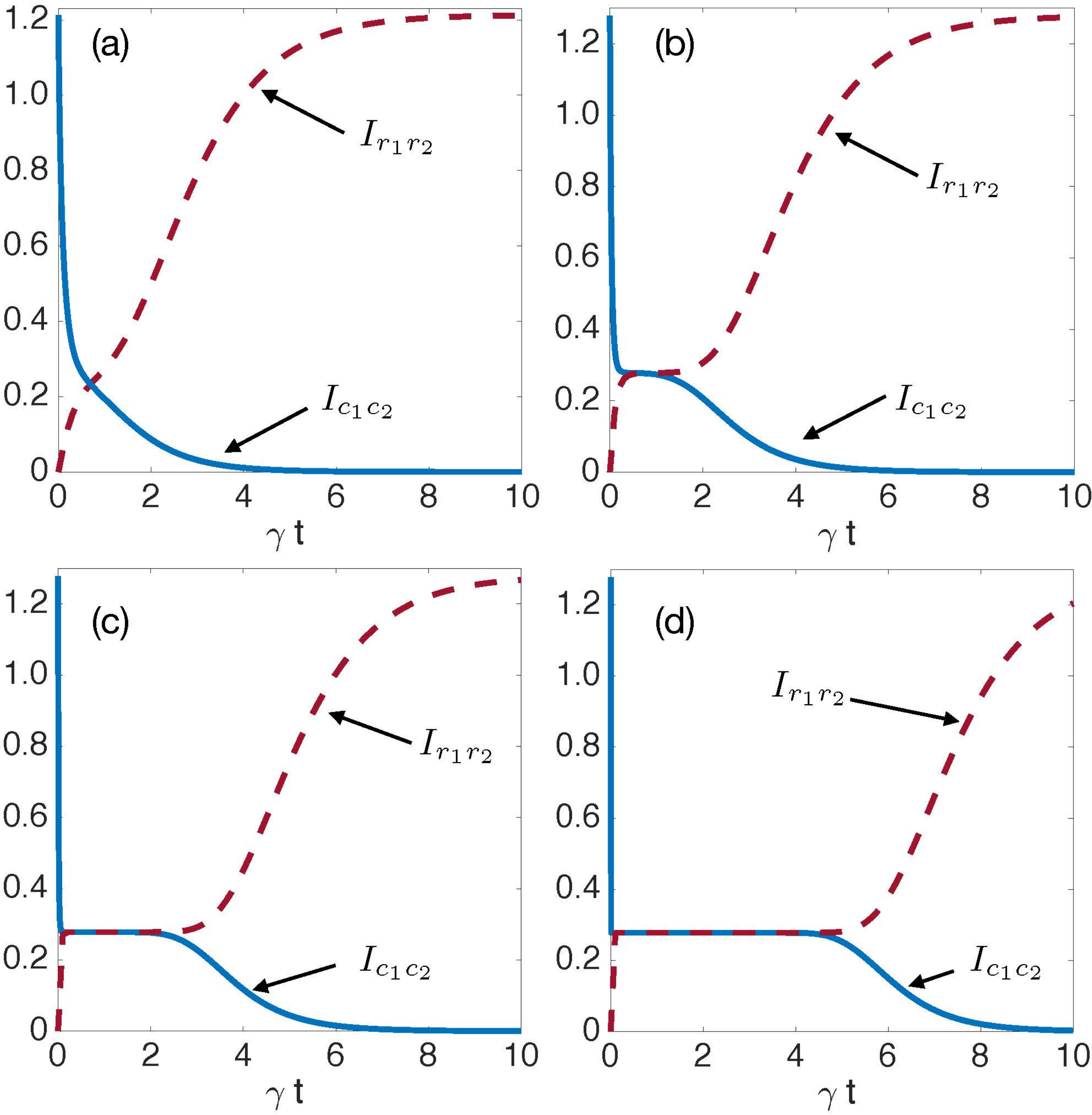}	
	\caption{ Evolution of quantum mutual information $I_{c_1 c_2 }$ (blue-solid line) and $I_{r_1 r_2}$ (red-dashed line) for (a) $\bar{n} = 1$, (b) $\bar{n} = 3$, (c) $\bar{n} = 10$ and (d) $\bar{n} = 100$. The initial state is~(\ref{rho0}) with $p=0.2$.}  
	\label{figmutual1}
\end{figure}

Now, we are in a position to study the dynamics of quantum and classical correlations. First, let us consider the dynamics of quantum mutual information $I_{c_1 c_2}$ and $I_{r_1 r_2}$ in the cavities and reservoirs subsystems. Their evolution is shown in Fig.~\ref{figmutual1} for different initial amplitudes $\alpha$ of the quantum field in the cavities. From the figures, it is evident that while mutual correlations in the cavities subsystem decrease, they increase in the reservoirs. The fact that $I_{c_1 c_2} (t=0) = I_{r_1 r_2}(t\rightarrow \infty)$ for any value of $\alpha$, shows us that there is a full transference of correlations from the cavities (blue solid line) to reservoirs (red dashed line). 

For cases where the value of $\bar{n}$ is higher, as in Figs.~\ref{figmutual1}(c) and~\ref{figmutual1}(d), we see that the dynamics of the quantum mutual information $I$ experiences a particular behavior: During a finite time, $I$ in both partitions is constant. 

The explanation of this behavior can be found in the dynamics of the cavities. From Eq.~(\ref{rhoc1c2}) it can be shown that during this time interval, the elements of the density matrix (\ref{rhoc1c2}) are: $r_{11}=r_{22}=r_{33}=r_{44} \simeq 1/4$ and $r_{14}=r_{41}=r_{23}=r_{32}= (2p-1)/4$. That is, for an initial state of form (\ref{rho0}) the system settles during this time interval in a decoherence-free subspace. Since we know the evolution of the density matrix, we can estimate the time interval $\Delta t$ during which the density matrix remains constant~\cite{lastra18}:
\begin{equation} \label{dt}
	\Delta t \simeq \frac{1}{\gamma}\ln{(\bar{n}-1)}
\end{equation}

This result is consistent with what we see in Fig.~\ref{figmutual1} which shows that $\Delta t$ increases with $\bar{n}=|\alpha|^2$.

\begin{figure}[t]
	\includegraphics[width=86mm]{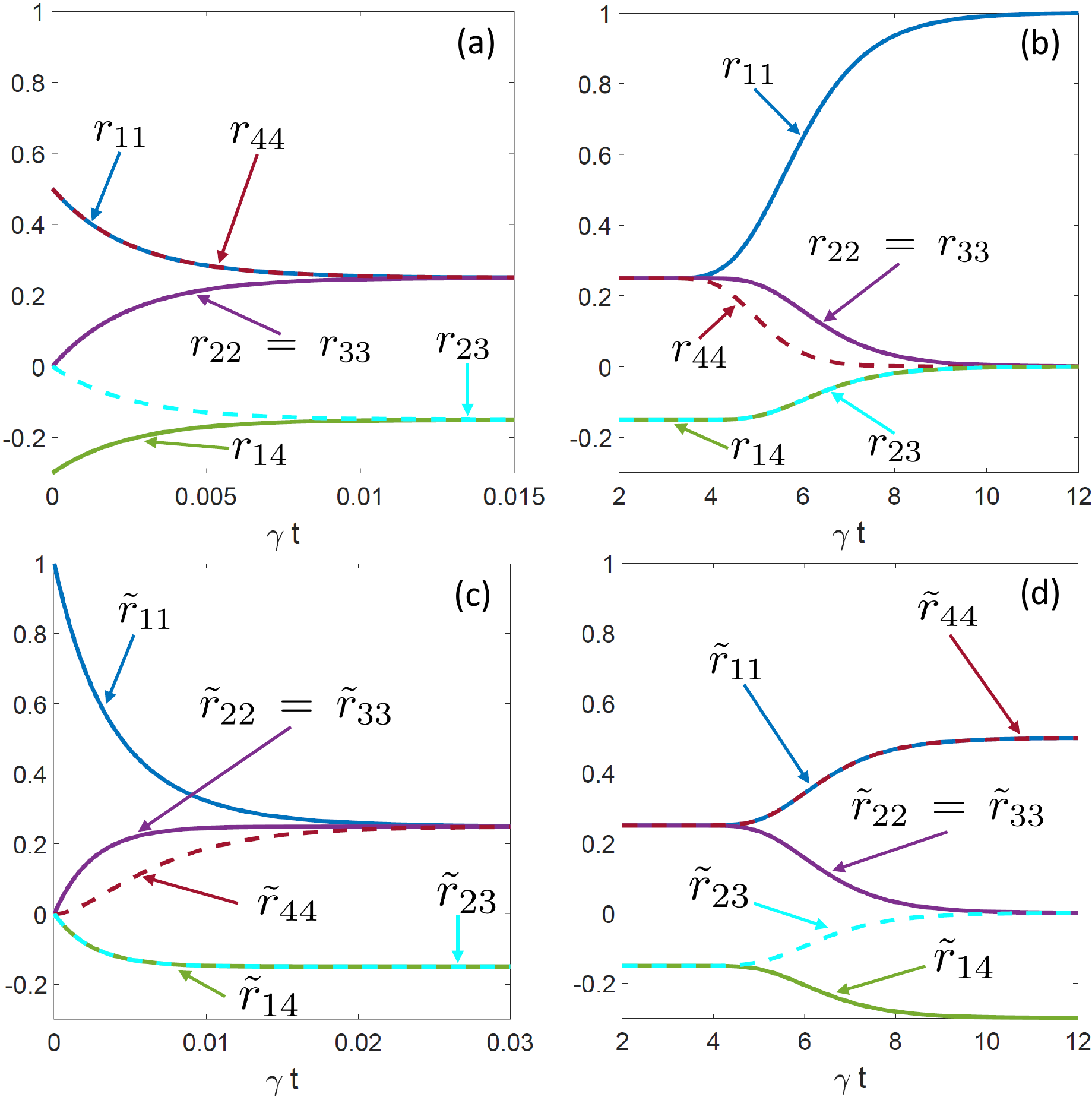}	
	\caption{Evolution of the matrix elements for two different time scales $\gamma t  \ll 1 $ and $\gamma t \geq 2$. Partition $c_1 \otimes c_2$: (a) and (b); Partition $r_1 \otimes r_2$: (c) and (d). The initial state is~(\ref{rho0}) with $p=0.2$.}  
	\label{ccrr}
\end{figure}

Interestingly, this behavior can also be observed in the reservoirs subspace $r_1 \otimes r_2$. In Fig.~\ref{ccrr}, the dynamics for the matrix elements for the cavity and reservoir systems are shown. It can be shown analytically that when the system reaches the decoherence-free space, the elements of the density matrix $\rho_{r_1 r_2}$ are: $\tilde{r}_{11}=\tilde{r}_{22}=\tilde{r}_{33}=\tilde{r}_{44} \simeq 1/4$ and $\tilde{r}_{14}=\tilde{r}_{41}=\tilde{r}_{23}=\tilde{r}_{32}= (2p-1)/4$, which resembles the behavior in the cavities system $\rho_{c_1,c_2}$ during a time $\Delta t$ given by $(\ref{dt})$. In the cavities, we say that the system reached a space free of decoherence. On the other hand, in the reservoirs partition, this time interval where the density matrix does not evolve, we can interpret it as stagnation in transferring the information from the cavities.

However, the similarities between the two partitions exist not only in the decoherence-free space: If we compare Figs.~\ref{ccrr}(a) and \ref{ccrr}(d), we see that the matrix elements for $\rho_{c_1 c_2}(t \rightarrow 0)$ are equal to those of the density matrix $\rho_{r_1 r_2}( t \rightarrow \infty)$. This is an analtical prove that the state of the cavities in $t = 0$ is completely mapped to the reservoirs when $t \rightarrow \infty$.

\begin{figure}[t]	
	\includegraphics[width=85mm]{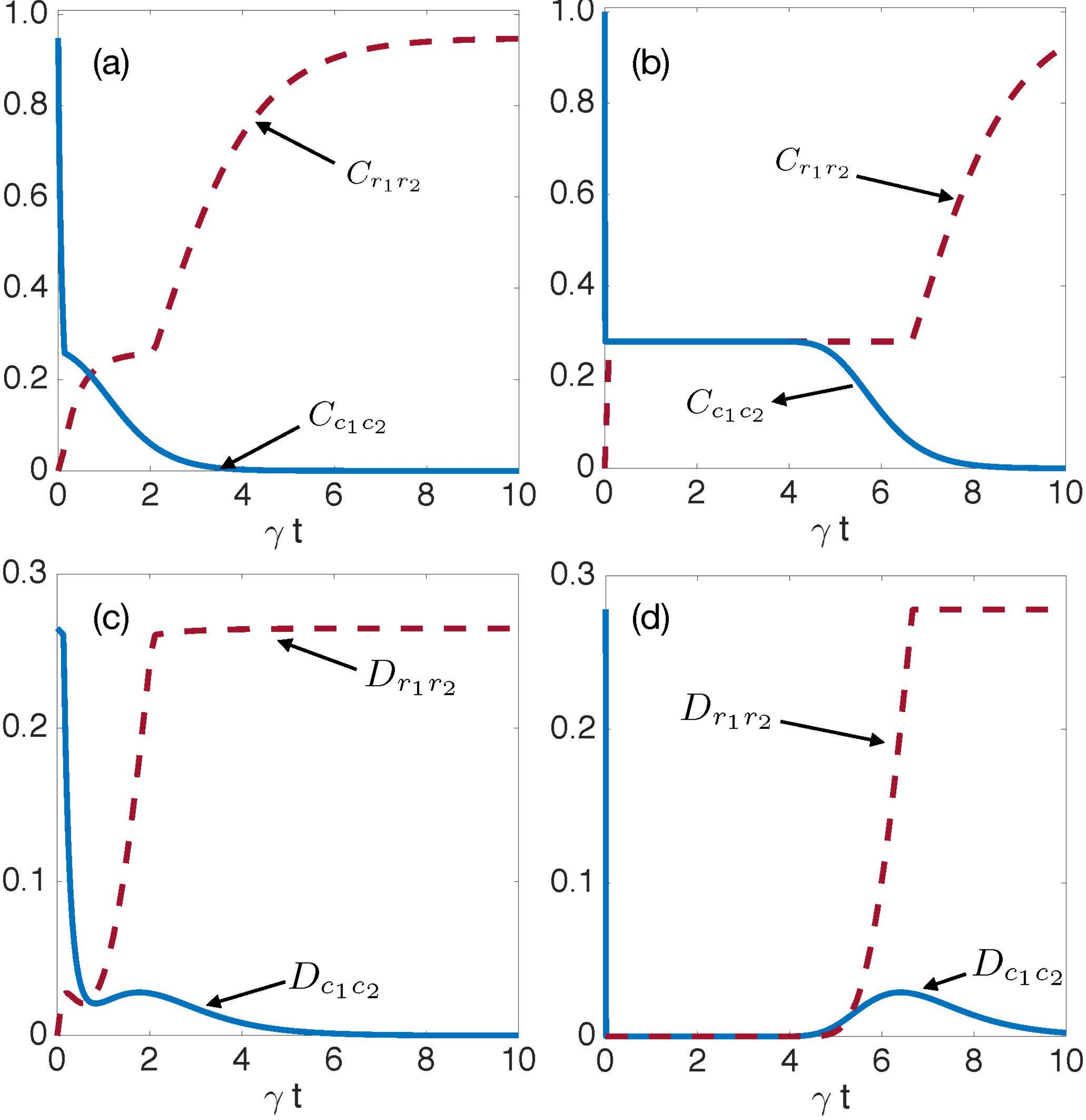}	
	\caption{ Evolution of classical correlations  $C_{c_1 c_2 }$ (blue-solid line) and $C_{r_1 r_2}$ (red-dashed line) for (a) $\bar{n} = 1$, (b) $\bar{n} = 100$, and the evolution of Discord $D_{c_1 c_2 }$ (blue-solid line) and $D_{r_1 r_2}$ (red-dashed line)  (c) $\bar{n} = 1$ and (d) $\bar{n} = 100$. The initial state is~(\ref{rho0}) with $p=0.2$.}  
	\label{figclassdiscord1}
\end{figure}

To further understand the evolution and transfer of correlations between cavities and reservoirs, we will study both classical and quantum correlations in detail. For example, in Fig.~\ref{figclassdiscord1}, we show both correlations for cavities and reservoirs. As for the mutual quantum information $I$, in this figure, we see a complete transfer of the classical and quantum correlations from the cavities to the reservoirs. Furthermore, in Figs.~\ref{figclassdiscord1}(b) and~\ref{figclassdiscord1}(d), we see that while the system populates the decoherence-free subspace, classical correlations are constant, and quantum correlation (discord) is zero. This result tells us that the system only has classical correlations in the subsystems of cavities and reservoirs in the decoherence-free subspace.

The freezing of quantum mutual information and, in particular, classical correlations are associated with the appearance of a pointer state basis~\cite{maniscalco,lastra2014,lastra17,lastra18}. 

\begin{figure}[t]
	\includegraphics[width=85mm]{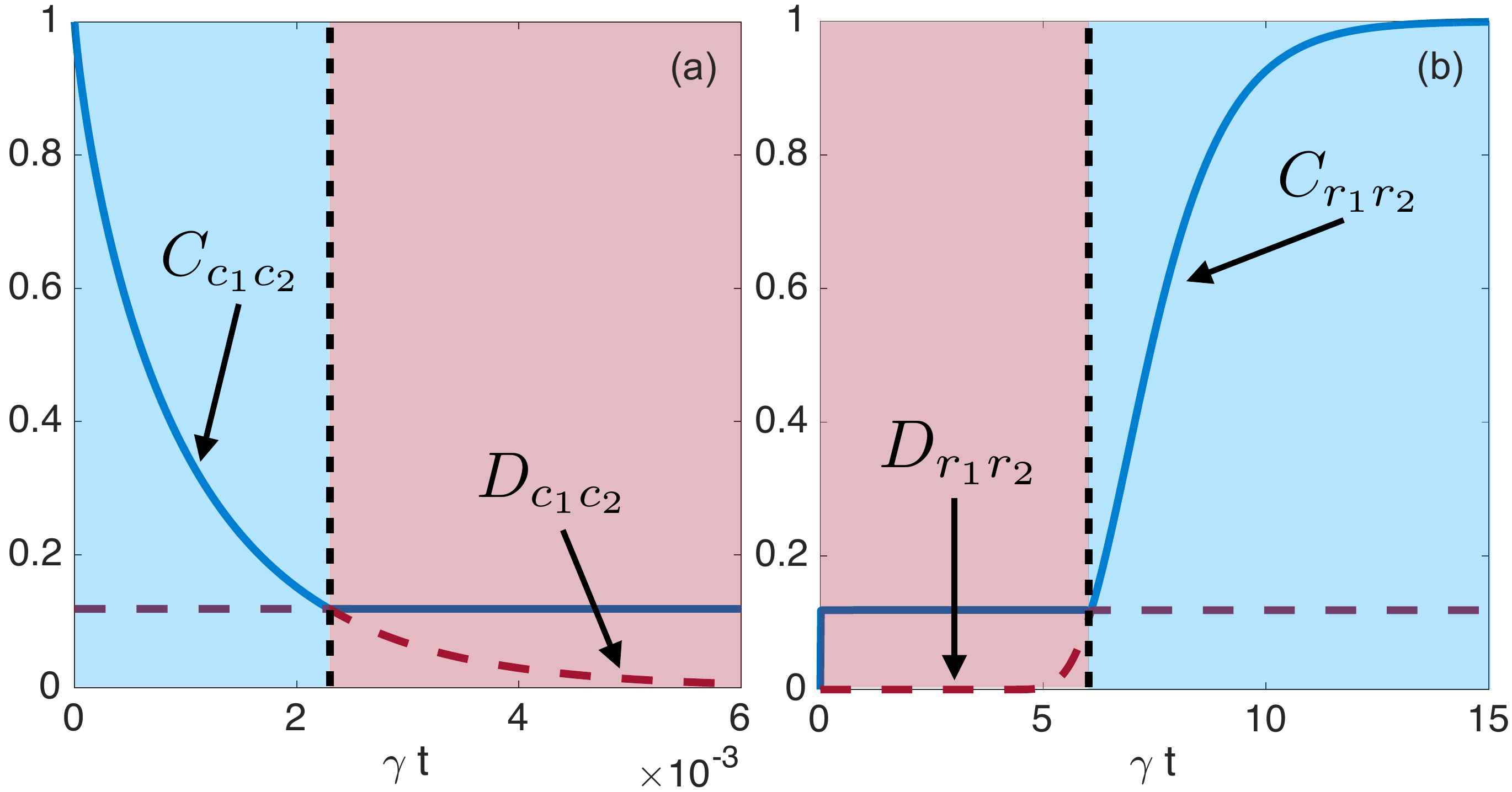}	
	\caption{Transference of quantum and classical correlations from cavities to the environment. (a) Classical correlations $C_{c_1 c_2}$ (blue solid-line) and discord (red dashed-line) $D_{c_1 c_2}$ in the cavities subsystem $(c_1 \otimes c_2)$. (b) Classical correlations $C_{r_1 r_2}$ (blue solid-line) and discord (red dashed-line) $D_{r_1 r_2}$ in the reservoirs subsystem $(r_1 \otimes r_2)$. The initial state is~(\ref{rho0}) with $\bar{n}= 100$ and $p=0.2$.Vertical black dashed-lines: (a) Time $\gamma t = \gamma t_c$; (b) $\gamma t = \gamma t_r$.}  
	\label{times}
\end{figure}

Both discord and classical correlations have a more complex dynamic compared to quantum mutual information. For example, in Fig.~\ref{times}(a) we see an interesting feature of the discord: For high values of $\bar{n}$ (here $\bar{n} = 100)$, it is observed that at a given time $t = t_c$ there is an {\it abrupt decay} of the discord $D_{c_1 c_2}$ in the cavities subspace. Simultaneously, the classical correlations $ C_ {c_1 c_2} $ also experience an abrupt decay. At this time, it holds that $D_{c_1 c_2} (t_ {c	}) = C_{c_1 c_2} (t_ {c})$. As of this moment, the classic correlations in the cavities freeze and the system reaches a metastable pointer state and then decays to zero. That is, it is completely transferred to the reservoirs.

On the other hand, in Fig.~\ref{times}(b) we see that there is a second relevant time $t = t_r$. This time $t_r$ marks a change of decoherence regime in the reservoirs. We see that before this time, the decoherence is quantum since $C_{r_1,r_2}$ is constant, but after $ t> t_r $ it is the discord $D_{r_1, r_2}$ that remains constant.

These two times $ t_c $ and $ t_r $ can be calculated analytically. Following the discussion before and after Eq.~(\ref{cca}), it is not difficult to show that $t_c$ also corresponds to the time where $C^{X}_{c_1 c_2} = C^{Z}_{c_1 c_2}$, while $t_r$ coincides with the time where $C^{X}_{r_1 r_2} = C^{Z}_{r_1 r_2}$. Solving these equations, for the partition $c_1 \otimes c_2$ we find that

\begin{equation}
t_c = -\frac{1}{\gamma} \ln{\left[1+\frac{1}{4 \bar{n}}\ln{\left(2p-1\right)}\right]}, \label{tc}
\end{equation}
while for the reservoir partitions $r_1 \otimes r_2$ we have that,
\begin{equation}
t_r = -\frac{1}{\gamma} \ln{\left[\frac{1}{4 \bar{n}}\ln{\left(\frac{1}{2p-1}\right)}\right]}.\label{tr}
\end{equation}

These equations show that both $t_c$ and $t_r$ depend on the intensity of the cavity fields and also on the parameter $p$ of the initial state of the system. It is also interesting to note that $t_c$ and $t_r$ are related through the following equation:
\begin{equation}
e^{-\gamma t_c}+e^{-\gamma t_{r}} = 1
\label{tctr}
\end{equation}

This relation between times of sudden transition resembles the one found for times of birth and sudden death of entanglement~\cite{lastra}.

In this Letter, we studied the correlations and their dynamics in a system of two mesoscopic quantum modes independently coupled to dissipative reservoirs. We have shown that total correlations are transferred from the cavities to the reservoirs regardless of the initial state. Also, we show that the total correlations' dynamics depend strongly on the initial amplitude of the quantum field in the cavities ($\bar{n}$). For example, with high values of $\bar{n}$, we see that the system visits a decoherence-free subspace during a finite time. Interestingly, this behavior also appears in the subspace of the reservoirs. In this time interval, where there is no loss of decoherence in the cavities' subsystem, information stops flowing into the reservoirs' subsystem. We showed analytically that the time interval depends on initial field amplitude.

When we focus on the quantum and classical correlations present in the system, we see that this decoherence-free space is associated with the freezing of classical correlations in both cavities and reservoirs, meaning that in both partitions, a metastable pointer state basis emerges.

\section{Acknowledgments }

Authors acknowledge financial support from DICYT Grant
No. 041931LC.


\begin{thebibliography}{99}
 	
\bibitem{Zeh70} H.D. Zeh, Found. Phys. \textbf{1}, 69 (1970).
\bibitem{zurek81}W.H. Zurek, Phys. Rev.
D \textbf{24}, 1516 (1981).
\bibitem{Kupsh03} E. Joos, H.D. Zeh, C. Kiefer, D.J.W. Giulini, J. Kupsch
and I.O. Stamatescu, Decoherence and the Appearance of
a Classical World in Quantum Theory, (Springer, Berlin,
2003), 2nd ed.
\bibitem{zurek03} W.H. Zurek, Decoherence, einselection, and the quantum origins of the classical, Rev. Mod. Phys. \textbf{75}, 715 (2003)
\bibitem{zurek97} J.R. Anglin, J.P. Paz, and W.H. Zurek, Phys. Rev. A \textbf{55}, 4041 (1997).
\bibitem{buhmann12} S. Scheel and S.Y. Buhmann, Phys. Rev. A \textbf{85}, 030101(R) (2012)
\bibitem{machnikowski06} P. Machnikowski, Phys. Ref. B \textbf{73}, 155109 (2006).
\bibitem{howie11} A. Howie, Ultramicroscopy \textbf{111}, 761 (2011). 	

\bibitem{zy} K. Z\"{y}czkowski, P. Horodecki, M. Horodecki, and R. Horodecki, Phys. Rev. A \textbf{65}, 012101 (2001).

\bibitem{dio} L. Di\'{o}si, Lect. Notes Phys. \textbf{622}, 157 (2003).

\bibitem{dod} P. J. Dodd and J. J. Halliwell, Phys. Rev. A \textbf{69}, 052105
(2004).

\bibitem{yu}Ting Yu and J.H. Eberly, Phys. Rev. Lett. \textbf{93}, 140404
(2004); 97, 140403 (2006).

\bibitem{san} M. F. Santos, P. Milman, L. Davidovich, and N. Zagury,
Phys. Rev. A \textbf{73}, 040305(R) (2006).

\bibitem{lan} B. P. Lanyon, M. Barbieri, M. P. Almeida, and A. G. White,
Phys. Rev. Lett.  \textbf{101}, 200501 (2008).

\bibitem{sha} A. Shabani and D. A. Lidar, Phys. Rev. Lett. \textbf{102}, 100402
(2009).

\bibitem{dat} A. Datta and S. Gharibian, Phys. Rev. A \textbf{79}, 042325
(2009).

\bibitem{pia} M. Piani et al., Phys. Rev. Lett. \textbf{77}, 250503 (2009).

\bibitem{wer} T. Werlang et al., Phys. Rev. A \textbf{80}, 024103 (2009).

\bibitem{maz1}J. Maziero et al., Phys. Rev. A \textbf{81}, 022116 (2010).

\bibitem{quantummutual} B. Groisman, S. Popescu, and A. Winter, Phys. Rev. A $\mathbf{72}$ 032317 (2005). 

\bibitem{oli}H. Ollivier and W. H. Zurek, Phys. Rev. Lett. \textbf{88}, 017901 (2001).

\bibitem{hen}L. Henderson and V. Vedral, J. Phys. A \textbf{34}, 6899 (2001).

\bibitem{opp}J. Oppenheim et al., Phys. Rev. Lett. \textbf{89}, 180402 (2002).

%\bibitem{Luo} S. Luo, Phys. Rev. A \textbf{77}, 022301 (2008).

\bibitem{Luo1} S. Luo, Phys. A. \textbf{77}, 042303 (2008).

\bibitem{mod} K. Modi, T. Paterek, W. Son, V. Vedral, and M. Williamson,
Phys. Rev. Lett. \textbf{104}, 080501 (2010).

\bibitem{maz} J. Maziero, L. C. C\'{e}leri, R. M. Serra and V. Vedral,Phys. Rev. A. \textbf{80}, 044102 (2009).

\bibitem{mazz} L. Mazzola, J. Piilo and S. Maniscalco, Phys. Rev. Lett. \textbf{104}, 200401 (2010).

\bibitem{xu} Jin-Shi Xu, Chuan-Feng Li, Cheng-Jie Zhang, Xiao-Ye Xu, Yong-Sheng Zhang, and Guang-Can Guo,Phys. A. \textbf{82}, 042328 (2010).

\bibitem{cor} M. F. Cornelio, O. Jimenez Faras, F. F. Fanchini, I. Frerot, G. H. Aguilar, M. O. Hor-Meyll, M. C. de Oliveira, S. P. Walborn, A. O. Caldeira, and P. H. Souto Ribeiro, Phys. Rev. Lett. \textbf{109}, 190402 (2012).

\bibitem{maniscalco} L. Mazzola, J. Piilo, and S. Maniscalco
Phys. Rev. Lett. {\bf 104}, 200401 (2010).

\bibitem{aaronson} B. Aaronson, R. Lo Franco, and G. Adesso, Phys. Rev. A \textbf{88}, 012120 (2013).

\bibitem{lastra2014} F. Lastra, C. E. L\'opez, S. A. Reyes, and S. Wallentowitz, Phys. Rev. A \textbf{90}, 062103 (2014).

\bibitem{lastra17} C.E. L\'opez and F. Lastra, Phys. Rev A \textbf{96}, 062112 (2017).

\bibitem{lastra18} F. Lastra. C.E. L\'opez and J.C. Retamal, Phys. Rev. A \textbf{97}, 042123 (2018).

\bibitem{zur} W.H. Zurek, Rev. Mod. Phys.  \textbf{75}, 715 (2003).


\bibitem{blu}R. Blume-Kohout and W.H. Zurek, Phys. Rev. A  \textbf{73},
062310, (2006); W. H. Zurek, Nat. Phys.  \textbf{5}, 181 (2009).

\bibitem{alm} M. P. Almeida, F. de Melo, M. Hor-Meyll, A. Salles, S. P. Walborn, P. H. S. Ribeiro, and L. Davidovich, Science  \textbf{316}, 579 (2007); O.J. Far\'{i}as, C.L. Latune, S.P. Walborn, L. Davidovich, and P. H. S. Ribeiro, Science  \textbf{324}, 1414 (2009).

\bibitem{bru} M. Brune, E. Hagley, J. Dreyer, X. Maöõtre, A. Maali, C. Wunderlich, J. Raimond, and S. Haroche, Phys. Rev. Lett.  \textbf{77}, 4887 (1996); C. Monroe, D. M. Meekhof, B. E. King, and D.J. Wineland, Science  \textbf{272}, 1131 (1996); J.P. Paz, S. Habib, and W. H. Zurek, Phys. Rev. D  \textbf{47}, 488 (1993).

\bibitem{lop} C. E. L\'{o}pez, G. Romero, F. Lastra, E. Solano, and J. C. Retamal,
Phys. Rev. Lett.  \textbf{101}, 080503 (2008).

\bibitem{lastra} F. Lastra, G. Romero, C.E. L\'{o}pez, N. Zagury, J.C. Retamal, Optics Communications, V 283, Issue 19, 3825 (2010).

\bibitem{xstate} M. Ali, \emph{et al.}, Phys. A. \textbf{81}, 042105 (2010).

\bibitem{CHOh2011} Qing Chen, Chengjie Zhang, Sixia Yu, X. X. Yi, and C. H. Oh
Phys. Rev. A \textbf{84}, 042313 (2011).  
		
\end{thebibliography}
\end{document}